\documentclass[12pt]{article}
\usepackage[english]{babel}
\usepackage[latin1]{inputenc}
\usepackage{color}

\usepackage[intlimits]{amsmath}
\usepackage{amssymb}

\usepackage{url}
\usepackage{graphicx}

\numberwithin{equation}{section}

\DeclareMathOperator{\BR}{BR}

\newcommand{\fex}{{\it e.g.}}

\newcommand{\GeV}{\,\text{GeV}}
\newcommand{\TeV}{\,\text{TeV}}

\newcommand{\s}{\,\text{s}}

\newcommand{\Mpc}{\,\text{Mpc}}

\newcommand{\be}{\begin{equation}}
\newcommand{\ee}{\end{equation}}
\newcommand{\bea}{\begin{eqnarray}}
\newcommand{\eea}{\end{eqnarray}}
\newcommand{\ppp}[1]{\begin{pmatrix} #1 \end{pmatrix}}
\def\stau{{\widetilde\tau}}
\def\lstau{{\widetilde\tau_1}}
\def\neut{{\chi_1^0}}
\def\hf{{X}} 

\newcommand{\OmTP}{\Omega_{3/2}^{{\rm th}}}

\newcommand{\OmNT}{\Omega_{3/2}^{\lstau}}
\newcommand{\OmSTP}{\Omega_{\hf}^{{\rm th}}}
\newcommand{\OmSNT}{\Omega_{\hf}^{\lstau}}

\begin{document}

\date{\mbox{ }}

\title{
{\normalsize
\vspace{-2.0cm} 
DESY 10-056\hfill\mbox{}\\
TUM-HEP 758/10\hfill\mbox{}\\
MIT-CTP-4145\hfill\mbox{}\\
}
\vspace{2.0cm}
\bf Supersymmetric Leptogenesis with a Light Hidden Sector\\[8mm]}

\author{Andrea De Simone$^a$, Mathias Garny$^b$, Alejandro Ibarra$^b$,
Christoph Weniger$^c$\\[2mm]
{\normalsize\it $^a$Center for Theoretical Physics, 
Massachusetts Institute of Technology,}\\[-0.05cm] 
{\normalsize\it Cambridge, Massachusetts, USA }\\[2mm]
{\normalsize\it $^b$Physik-Department T30d, Technische Universit\"at
M\"unchen,}\\[-0.05cm]
{\it\normalsize James-Franck-Stra\ss{}e, 85748 Garching, Germany}\\[2mm]
{\normalsize\it $^c$Deutsches Elektronen-Synchrotron DESY, Hamburg}\\[-0.05cm]
{\it\normalsize Notkestra\ss{}e 85, 22603 Hamburg, Germany}
}
\maketitle

\thispagestyle{empty}

\begin{abstract}
\noindent

Supersymmetric scenarios incorporating thermal leptogenesis as the origin of
the observed matter-antimatter asymmetry generically predict abundances of the
primordial elements which are in conflict with observations.
In this paper we propose a simple way to circumvent this tension and
accommodate naturally thermal leptogenesis and primordial nucleosynthesis.
We postulate the existence of a light hidden sector, coupled very  weakly to
the Minimal Supersymmetric Standard Model, which opens up new decay channels
for the next-to-lightest supersymmetric particle, thus diluting its abundance
during nucleosynthesis.
We present a general model-independent analysis of this mechanism as well as
two concrete realizations, and describe the relevant cosmological and
astrophysical bounds and implications for this dark matter scenario.
Possible experimental signatures at colliders and in cosmic-ray observations
are also discussed.

\end{abstract}

\newpage

\section{Introduction}
\label{Introduction}

Extending the Minimal Supersymmetric Standard Model (MSSM) with three heavy
right-handed neutrino superfields is one of the best motivated scenarios for
physics beyond the Standard Model.  The decoupling of the heavy degrees of
freedom induces at low energies, after the electroweak symmetry breaking, tiny
neutrino masses suppressed by the large right-handed neutrino masses; this is
the renown see-saw mechanism~\cite{seesaw}. Furthermore, the out of
equilibrium decay of the lightest right-handed (s)neutrinos in the early
Universe could have generated the observed matter-antimatter asymmetry through
the mechanism of leptogenesis~\cite{Fukugita:1986hr}. Successful thermal
leptogenesis requires, though, a rather large mass scale for the new
particles, $M\gtrsim 10^9\GeV$~\cite{Davidson:2002qv}, which could destabilize
the electroweak scale.  Supersymmetry guarantees that the large quadratic
quantum corrections to the Higgs mass introduced by the right-handed
sneutrinos exactly cancel with the ones introduced by the right-handed
neutrinos, thus avoiding the severe hierarchy problem of the
non-supersymmetric version of the leptogenesis mechanism.

It is remarkable that this simple scenario can simultaneously address two of
the most severe limitations of the Standard Model, namely the existence of
non-vanishing neutrino masses and the origin of the observed matter-antimatter
asymmetry in our Universe (as well as the above mentioned hierarchy problem).
This appealing scenario is nevertheless not exempt of problems.  Gravitinos
are very efficiently produced in the very hot plasma necessary to generate the
observed baryon asymmetry through the mechanism of thermal leptogenesis, the
relic density being~\cite{Bolz:2000fu}:
\begin{equation}
  \OmTP h^2\simeq 0.27
  \left(\frac{T_R}{10^{9}\GeV}\right)
  \left(\frac{10\GeV}{m_{3/2}}\right)
  \left(\frac{m_{\widetilde g}}{1\TeV}\right)^2\;,
  \label{eqn:OmegaGravTP}
\end{equation}
where $m_{3/2}$ is the gravitino mass, $m_{\widetilde g}$ is the gluino mass
and $T_R$ is the reheating temperature of the Universe.  If the gravitino is
heavier than the lightest neutralino, it decays during or after Big Bang
Nucleosynthesis (BBN). Being the gravitinos so abundant at the time of
formation of the primordial elements, {\it cf.} Eq.~\eqref{eqn:OmegaGravTP},
the large hadronic energy injected into the primeval plasma destroys the
successful predictions of the Standard BBN scenario unless the gravitino mass
is larger than $\sim 10^4\GeV$~\cite{Kawasaki:2004qu, Cyburt:2009pg}.

On the other hand, if the gravitino is lighter than any observable
supersymmetric particle, it constitutes a natural candidate for the cold dark
matter of the Universe provided it is stable at cosmological scales.  Namely,
following Eq.~\eqref{eqn:OmegaGravTP}, the dark matter relic density inferred
by WMAP for the $\Lambda$CDM model, $\Omega_{\rm CDM} h^2\simeq
0.11$~\cite{Komatsu:2010fb}, can be reproduced for the range of reheating
temperatures required by thermal leptogenesis, $T_R\gtrsim 10^9\GeV$ and for
typical gluino masses, $m_{\widetilde g}\sim 1\TeV$, provided the gravitino
mass is larger than $\sim 10\GeV$.

This attractive scenario is in general in conflict with the observed
abundances of primordial elements.  If R-parity is conserved, the Lightest
Observable Supersymmetric Particle (LOSP) can only decay into Standard Model
particles and the gravitino with a decay rate suppressed by the Planck scale,
the lifetime being:
\be
\tau_{\rm LOSP} \simeq 3\,{\rm days}
\left(\frac{m_{3/2}}{10\GeV}\right)^2
\left(\frac{250\GeV}{m_{\rm LOSP}}\right)^{5}\;.
\label{LOSP-lifetime}
\ee
Therefore, the LOSP is typically present during or after BBN, jeopardizing the
successful predictions of the standard nucleosynthesis scenario. This is in
fact the case for the most likely candidates for the LOSP: the lightest
neutralino and the right-handed stau.  More precisely, when the LOSP is the
neutralino, the hadrons produced in the neutralino decays typically dissociate
the primordial elements~\cite{Kawasaki:2004qu, Jedamzik:2006xz}, yielding
abundances in conflict with observations. On the other hand, when the LOSP is
a charged particle, $X^-$, the formation of the bound state $(^4{\rm
He}\,X^-)$ catalyzes the production of $^6$Li~\cite{Pospelov:2006sc} leading
to an abundance of $^6$Li in stark conflict with
observations~\cite{catalyzedBBN} (for a recent review about BBN constraints
see Ref.~\cite{Jedamzik:2009uy}).

Different solutions have been proposed to this problem. For instance, in some
specific supersymmetric models the LOSP can be a
sneutrino~\cite{Kanzaki:2006hm} or a stop~\cite{DiazCruz:2007fc}, whose late
decays do not substantially affect the predictions of BBN.  For neutralino or
stau LOSP a possible solution consists in introducing a small amount of
$R$-parity violation, so that the LOSP decays into two Standard Model
particles before the onset of BBN, thus avoiding the BBN constraints
altogether~\cite{Buchmuller:2007ui}. Maintaining the requirement of R-parity
conservation, other solutions are to assume a large left-right mixing of the
stau mass eigenstates~\cite{left-right}, a LOSP mass that is nearly degenerate
with the gravitino mass~\cite{Boubekeur:2010nt}, or to assume some amount of
entropy production after LOSP decoupling, which dilutes the LOSP
abundance~\cite{Pradler:2006hh}.

In this paper we would like to propose a scenario which yields a thermal
history of the Universe consistent with supersymmetric dark matter and with
baryogenesis through thermal leptogenesis, without altering the successful
predictions of the Standard BBN scenario.\footnote{See also
Ref.~\cite{Asaka:2000ew} for a related proposal with light axinos, however
without supersymmetric dark matter.} We will assume the existence of a hidden
sector fermion, $\hf$, lighter than the LOSP. Thus, new decay channels are
possible for the LOSP, for instance, when the LOSP is the lightest stau or the
lightest neutralino,
\begin{eqnarray}
  \nonumber
  \lstau&\to&\tau \hf\;, \\
  \neut&\to&(Z^0, \gamma, h^0, f\bar f) \hf \;.
  \nonumber
\end{eqnarray}
If these decays are fast enough, the density of LOSPs at the time of
nucleosynthesis can be significantly reduced and thus the successful
predictions of the standard BBN scenario will not be jeopardized.  If the
gravitino is the LSP, the hidden sector fermion will eventually decay into the
gravitino and other particles, hidden or observable.  The late decays into
gravitinos and hidden sector particles, if kinematically possible, may disrupt
the abundances of {\it hidden sector} primordial  nuclei, but not the
abundances of the observed primordial nuclei. On the other hand, the decays
into gravitinos and observable particles occur at a rate much larger than the
age of the Universe, as we will show, thus not affecting primordial
nucleosynthesis.  Lastly, if the gravitino is not the LSP it will decay into
hidden sector particles. Again, these decays may disrupt the primordial
abundances of hidden sector nuclei, but will not have any impact on the
standard BBN predictions.  This mechanism is sketched in Fig.1, for the case
where the gravitino is the LSP (left panel) and for the case where the
gravitino can decay into hidden sector particles (right panel).

The couplings of the hidden sector fermion to the MSSM particles are subject
to a series of constraints which will be discussed in Section 2. In Section 3
we present two concrete models where the mechanism sketched above can be
implemented. In Section 4 we will comment on possible signatures of this
scenario at colliders or at cosmic-ray observations. Lastly, in Section 5, we
will present our conclusions.

\section{Hidden sector couplings to the MSSM}
We will consider in this paper a scenario where the MSSM particle content is
extended with a light superfield (chiral or vector), which is a singlet under
the Standard Model gauge group. We will further assume that the fermionic
component of this superfield couples to the LOSP and its Standard Model
counterpart via a tiny Yukawa coupling, hence we will refer to this fermion as
``hidden fermion''. In this section we will carefully discuss the implications
for leptogenesis of the existence of such hidden fermion, as well as the
constraints on this scenario from BBN and from structure formation. Let us
first discuss the case of stau LOSP and later on the case of neutralino LOSP.

\subsection{Stau LOSP}
The interaction Lagrangian between the hidden fermion and the lightest stau,
$\lstau$, is given by the renormalizable term
\begin{equation}
  -{\cal L}=\lambda_\lstau \bar \hf\,\tau\,\lstau +{\rm h.c.}\;.
  \label{eqn:Lagrangian1}
\end{equation}
Then the stau can decay either $\lstau\rightarrow \psi_{3/2}\,\tau$ or
$\lstau\rightarrow \hf\,\tau$ with decay rates:
\begin{equation}
  \Gamma_{\lstau \rightarrow \psi_{3/2} \tau} \simeq  {1\over
  48\pi}{m_{\lstau}^5\over m_{3/2}^2 m_P^2}\left(1-{m_{3/2}^2\over
  m_{\lstau}^2}\right)^{4}\;, 
  \label{eq:rate-into-grav}
\end{equation}
\begin{equation}
  \Gamma_{\lstau \rightarrow \hf \tau}
  \simeq \frac{ |\lambda_{\lstau}|^2 m_{\lstau} }{8\pi} \left( 1 -
  \frac{m_\hf^2}{m_{\lstau}^2} \right)^2 \,,
\label{eq:rate-into-X}
\end{equation}
where the tau mass has been neglected. If the coupling $\lambda_\lstau$ is
large enough, the stau will decay before it can form bound states with $^4$He,
thus preventing the catalytic production of $^6$Li. More concretely, the
requirement that the lightest stau decays before $\simeq 2\times 10^3$ s
\cite{Pospelov:2006sc, catalyzedBBN} into $\hf\tau$, implies the following
lower bound on the coupling $\hf$--$\tau$--$\lstau$:
\begin{equation}
  |\lambda_\lstau| \gtrsim 5 \times 10^{-15} \left({250 \GeV\over
  m_{\lstau}}\right)^{1/2}\left(1-{m_\hf^2\over m_{\lstau}^2}\right)^{-1} \,.
  \label{lower}
\end{equation}
If this condition is satisfied, the lightest stau will by itself not play any
role during BBN.\medskip

A necessary requirement for the viability of this mechanism is that
cosmological constraints, namely from thermal overproduction of the hidden
fermion $X$ and from structure formation, are satisfied. Different scenarios
can arise depending on whether the LSP is the gravitino or the hidden fermion,
and on whether the NLSP is stable on cosmological time-scales or not. Below we
will discuss each case separately.
 
\subsubsection{Cosmologically stable gravitino and hidden fermion}
\label{sec:Stable}
\begin{figure}[t]
  \begin{center}
    \includegraphics{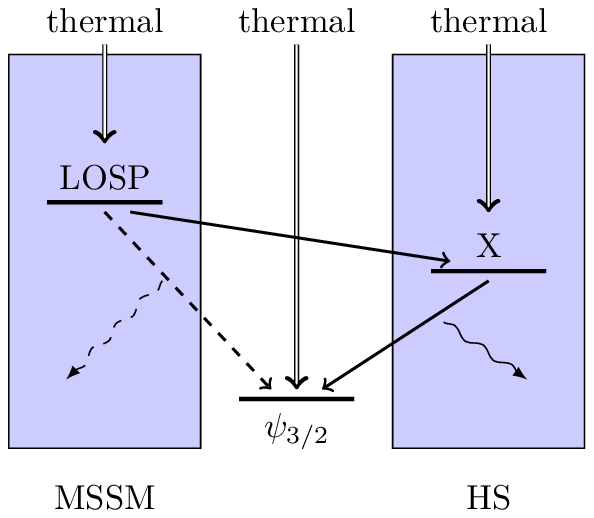}\hspace{2cm}
    \includegraphics{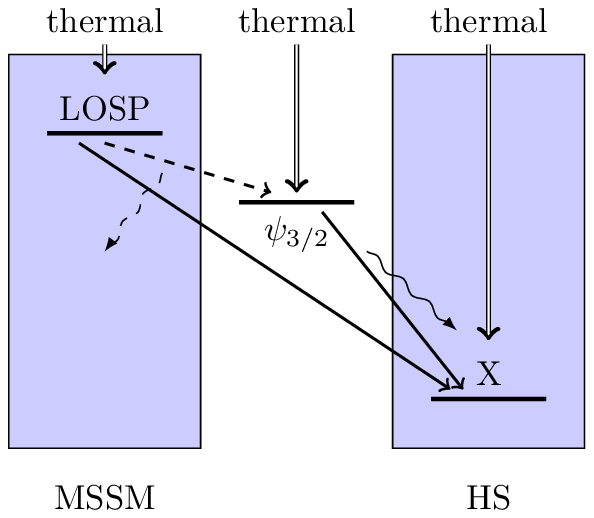}
  \end{center}
  \caption{Sketch of our proposed mechanism, for the cases where the LSP is
  the gravitino (\textit{left panel}) or a hidden fermion (\textit{right
  panel}).  The Lightest Observable Supersymmetric Particle (LOSP), the
  gravitino and the hidden fermion can be produced thermally or non-thermally
  through the decays of heavier particles; slow decays are indicated with
  dashed arrows and fast decays with solid arrows, whereas the other decay
  products are indicated with wiggled lines. If thermal leptogenesis is the
  correct mechanism to explain the observed matter-antimatter asymmetry, the
  LOSP decay into the gravitino occurs during or after the time of primordial
  nucleosynthesis, altering the predictions of the Standard BBN scenario. If
  this is the only decay channel of the LOSP the impact is usually dramatic,
  yielding abundances in conflict with observations.  However, if the LOSP
  coupling to the hidden fermion is much larger than the coupling to the
  gravitino, this decay can occur before the onset of the nucleosynthesis
  reactions, thus avoiding altogether any possible effect of the LOSP on
  nucleosynthesis.  Eventually the hidden fermion will decay into the
  gravitino and other hidden sector particles (\textit{left panel}) or vice
  versa (\textit{right panel}). Nevertheless, these decays do not alter the
  abundances of primordial elements in our observable sector.}
  \label{fig:sketch1}
\end{figure}

If the hidden fermion is the lightest particle in the hidden sector, its decay
channels into a gravitino and other hidden sector particles are kinematically
forbidden (note that in this case the hidden fermion still can decay into a
gravitino and Standard Model particles, {\it e.g.} $\hf\rightarrow \psi_{3/2}
\tau^+ \tau^-$, $\hf\rightarrow \psi_{3/2} \gamma$, with lifetimes which can
be larger than the age of the Universe, as will be discussed in Section 4). If
this is the case, two particle species contribute to the dark matter, namely
gravitinos and hidden fermions, each of them having a thermal component and a
non-thermal component:
\begin{equation}
  \Omega_\text{dm}  = \OmTP  + \OmNT  + \OmSTP  + \OmSNT  \,.
\end{equation}
Here, $\OmNT$ and $\OmSNT$ are the non-thermal contributions to the gravitino
and hidden fermion relic density, respectively, which are given by
\begin{eqnarray}
  \OmNT           & = & \frac{m_{3/2}}{m_{\lstau}} \,
  \mbox{BR}(\lstau\rightarrow\psi_{3/2}\tau ) \, \Omega_{\lstau}^{\rm th}
  \,,\\ \OmSNT  & = & \frac{m_{\hf}}{m_{\lstau}} \,
  \mbox{BR}(\lstau\rightarrow \hf\tau ) \, \Omega_{\lstau}^{\rm th} \,,
  \label{OmSNTP}
\end{eqnarray}
where $\Omega_{\lstau}^{\rm th}$ is the stau thermal abundance 
\begin{equation}\label{OmStau}
  \Omega_{\lstau}^{\rm th}h^2 \simeq 2 \times 10^{-3}
  \left(\frac{m_{\lstau}}{100\GeV}\right)^2 \,,
\end{equation}
corresponding to a yield $Y_{\lstau} \simeq 7 \times 10^{-14} \left(m_{\lstau}
/ 100\GeV\right)$ \cite{Asaka:2000zh}.  Furthermore, $\OmTP$ is the
contribution to the total dark matter density from thermally produced
gravitinos, Eq.~\eqref{eqn:OmegaGravTP}. If the hidden fermion couples to the
observable sector through a renormalizable coupling, the thermal production
proceeds dominantly via the decay of thermally produced staus
at temperatures $T \sim m_\lstau$. The
corresponding hidden fermion relic abundance $\OmSTP$ is given
by~\cite{Hall:2009bx}
\bea
\OmSTP h^2 & \simeq &
{1.09 \times 10^{27}\over g_*^{3/2}}{2m_{\hf}\Gamma_{\lstau \rightarrow \tau\hf
}\over m_{\lstau}^2}(1+\delta) \nonumber\\
& \simeq & 8.6\times10^{22}(1+\delta)
|\lambda_{\lstau}|^2 \left( \frac{100}{g_\ast} \right)^{3/2}
\left( \frac{m_\hf}{m_{\lstau}} \right)\left(
1-\frac{m_\hf^2}{m_{\lstau}^2} \right)^2\;.
\label{eqn:Xproduction1}
\eea
Here, $g_\ast\approx100$ denotes the effective number of degrees of freedom at
temperature $T\sim m_{\lstau}$, and $\delta$ parameterizes the potential
enhancement of the hidden fermion abundance due to possible additional
couplings of the hidden fermion to other MSSM particles,
\textit{cf.}~Eq.~\eqref{eqn:DefDelta} below.  Here we set for simplicity
$\delta=0$.

In order to sufficiently reduce the number density of staus at the time of BBN
it is necessary that $\text{BR}(\lstau\to\tau \hf)\simeq1$, and therefore,
$\Omega_{3/2}^{\lstau}\simeq0$. Requiring that the total dark matter density
does not exceed the measured value by WMAP implies then
\begin{align}
  \Omega^\text{th}_{3/2} + \Omega_\hf^\text{th}
  +\frac{m_X}{m_{\lstau}}\Omega_{\lstau}^\text{th}
  \lesssim 0.11 h^{-2} \;.
  \label{eqn:bounds1}
\end{align}
As long as the thermal production of $\hf$ and staus is small (which also
implies a small coupling $\lambda_{\lstau}$) this bound reduces to the
standard overproduction constraint on thermally produced gravitinos, $\OmTP
h^2 \lesssim 0.11$, which by its own allows high reheating temperatures
$T_R\gtrsim10^9\GeV$, as required by leptogenesis, provided the gravitino mass
is large enough, see Eq.~\eqref{eqn:OmegaGravTP}.  On the other hand, in the
regime where the production of $\hf$ is sizeable,
Eqs.~\eqref{eqn:Xproduction1} and \eqref{eqn:bounds1} combine to a bound on
the coupling 
\begin{equation}
  |\lambda_{\lstau}|
  \lesssim 10^{-12} \left(\frac{m_\lstau}{m_\hf }\right)^{1/2}
  \left(1-\frac{m_\hf^2}{m_\lstau^2}\right)^{-1}\;,
  \label{eq:lambda-upperbound}
\end{equation}
which is independent of the gravitino mass. 

\begin{figure}[t]
  \begin{center}
    \includegraphics[width=0.6\linewidth]{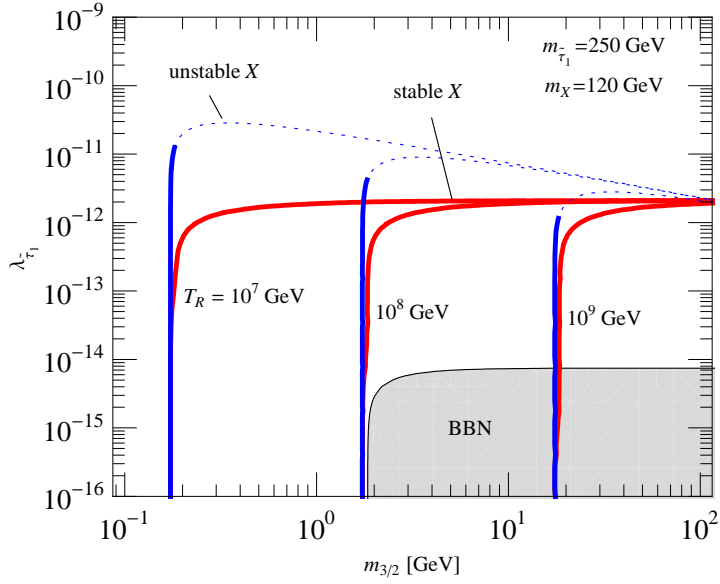}
  \end{center}
  \caption{Summary of constraints on stau LOSP scenario with \textit{gravitino
  LSP}, as derived from Eq.~\eqref{eqn:Lagrangian1}, as function of
  the stau-tau-hidden fermion Yukawa-coupling $\lambda_{\lstau}$ and gravitino mass $m_{3/2}$. The masses
  of $\hf$ and $\lstau$ are fixed as indicated. The \textit{red} and
  \textit{blue lines} show the values of $m_{3/2}$ and $\lambda_{\lstau}$ that
  yield the correct total relic abundance for different reheating temperatures
  $T_R=10^7$--$10^9\GeV$, assuming that $X$ is stable or unstable,
  respectively (the gluino mass has been set to $m_{\tilde g}=800\GeV$).
  Furthermore, the \textit{dashed} part of the blue lines is excluded by
  constraints on mixed warm/cold dark matter as discussed in the text, and the
  gray region is excluded by $^6$Li overproduction during BBN. It is clear
  that for vanishing coupling to the hidden sector, reheating temperatures
  around $10^9\GeV$, as required by thermal leptogenesis, are in conflict with
  BBN whereas an allowed window opens up for non-zero $\lambda_{\lstau}$.}
  \label{fig:summaryI}
\end{figure}

The results are illustrated by the red lines in Fig.~\ref{fig:summaryI} for
the case of gravitino LSPs and in Fig.~\ref{fig:summaryII} for the case of
hidden fermion LSPs. The lines show, for different reheating temperatures and
different masses of the hidden fermion, the value of the coupling
$\lambda_\lstau$ as a function of the gravitino mass from the requirement that
the {\it total} dark matter density is equal to the value inferred by the WMAP
collaboration. In these plots, it was assumed for definiteness
$m_\lstau=250\GeV$. On the other hand, the shaded regions correspond to
choices of parameters where the stau decays with a lifetime longer than
$2\times 10^3\s$, thus leading to $^6$Li overproduction.  As apparent from the
plots, in both cases of hidden fermion LSP and gravitino LSP, there is a
fairly wide region of the parameter space where the reheating temperature can
be large enough to allow thermal leptogenesis while preserving the successful
predictions of the standard BBN scenario.

\subsubsection{Unstable hidden fermion}
If the gravitino is the LSP, and if kinematically allowed, the hidden fermion
$\hf$ decays into gravitinos and hidden sector particles well before
matter-radiation equality.  Note that the particles produced in the decay
interact very weakly with the particles in the observable sector, therefore
the late decays of the hidden fermion do not modify the abundances of
primordial elements in the observable sector.  In this scenario, the dark
matter consists of thermally produced gravitinos, with a relic density given
by Eq.~\eqref{eqn:OmegaGravTP}, and non-thermally produced gravitinos, coming
from the late decay of hidden fermions $X$ and staus $\lstau$. The dark matter
abundance is then given by
\begin{align}
  \Omega_\text{dm}=\Omega_{3/2}^\text{th} +   \underbrace{
  \Omega_{3/2}^{\lstau} +
  \frac{m_{3/2}}{m_\hf}\left(\Omega_\hf^\text{th} +
  \Omega_X^{\lstau}\right)}_{=\Omega_\text{WDM}}\;,
  \label{eqn:abu2}
\end{align}
where we assumed for simplicity that the hidden-sector particles produced in
the decay of $X$ are massless or very light and hence contribute negligibly to
the relic abundance.\footnote{Otherwise they would contribute to the warm dark
matter component and they could cause dangerous late decays into Standard
Model particles.} The component coming from the late decay of $\hf$, as well
as the small fraction of gravitinos produced directly in $\lstau$ decays, will
typically act as warm dark matter (WDM), with free-streaming lengths
$\lambda_\text{FS}\gtrsim5\Mpc$ (\textit{cf.}~Fig.~4~in
Ref.~\cite{Ibarra:2008kn}).\footnote{Note that this could also be relevant for
$\Omega_\hf^{\lstau}$ in the above case where gravitinos and hidden fermions
are stable, since for small hidden gaugino masses the free-streaming length
becomes large. However, in the example shown in Fig.~\ref{fig:summaryI}, the
red lines feature always free-streaming lengths below $\lambda_\text{FS}\ll
0.5\Mpc$ in the region where $\lambda_{\lstau}\gtrsim 10^{-14}$.
Furthermore, since the stau yield is small, the impact of this component on
mixed warm/cold dark matter bounds is negligible even if the free-streaming lengths
are large.} Here, the free-streaming length of a particle is defined as the
distance the particle has traveled between its production and the onset of
structure formation. It is given by
\begin{align}
  \lambda_\text{FS}=\int^{z_p}_{3000} dz\frac{v(z)}{H(z)}\;,
  \label{eqn:defFS}
\end{align}
where $z_p$ denotes the red-shift at which the particle is produced and is a
function of the lifetime of the parent particle, $z\sim3000$ is the redshift
at matter-radiation equality, $v(z)$ is the particle's velocity and $H(z)$
denotes the Hubble parameter as function of redshift, see
\textit{e.g.}~Ref.~\cite{Kolb:1990vq}. 

Observations of the power spectrum of high-redshift Hydrogen clouds via the
Lyman-$\alpha$ forest~\cite{McDonald:2004eu} imply the upper bound
$\lambda_\text{FS}\lesssim 0.5\Mpc$~\cite{Strigari:2006jf}, when all dark
matter components have comparable free-streaming lengths.  This bound relaxes
if a large fraction of the dark matter is cold and just a small fraction of it
is warm. Bounds on the fraction $f$ of the dark matter density that is allowed
to be warm with a free-streaming length above $0.5\Mpc$, were discussed in
Ref.~\cite{Boyarsky:2008xj} in the context of sterile neutrinos. There, using
Lyman-$\alpha$ data~\cite{Viel:2004bf} and WMAP5 results, $2\sigma$-bounds
around $f\lesssim0.05$ were found for a warm component with free-streaming
lengths around $\mathcal{O}(10\Mpc)$, corresponding to $\mathcal{O}(1\
\text{km}/\s)$ thermal velocities.  

For definiteness we will take $f=0.05$ throughout this paper, which implies,
following Eq.~\eqref{eqn:abu2}, the requirement of a small thermal abundance
of $\lstau$ and $\hf$. Allowing a fraction $f$ of dark matter to be warm, and
provided that $\Omega_{3/2}^{\lstau}\simeq0$, gives then
\begin{align}
  \frac{m_{3/2}}{m_X}\left(\Omega_\hf^\text{th} +
  \Omega_\hf^{\lstau}\right)
  \lesssim f\ 0.11\ h^{-2}\;.
  \label{eqn:WDMbounds}
\end{align}
In addition to this constraint, the viability of the present scenario requires
that both, the thermal gravitino abundance and the thermal hidden fermion
abundance, do not exceed the total dark matter density, namely
$\Omega_{3/2}^\text{th}+ (m_{3/2}/m_\hf)\Omega^\text{th}_\hf \lesssim 0.11\
h^{-2}$, \textit{cf.} Eq.~\eqref{eqn:Xproduction1}.

The impact of all these constraints on the parameter space of the scenario is
shown in Fig.~\ref{fig:summaryI} as blue lines. The dashed part of the lines
is excluded from the constraints on mixed warm/cold dark matter,
Eq.~\eqref{eqn:WDMbounds}. Compared to the scenario where the hidden fermion
is stable, now larger values of the coupling $\lambda_\lstau$ are allowed,
since the hidden fermion does not directly contribute to the total dark matter
abundance any more.

\subsubsection{Unstable gravitino}
\begin{figure}[t]
  \begin{center}
    \includegraphics[width=0.6\linewidth]{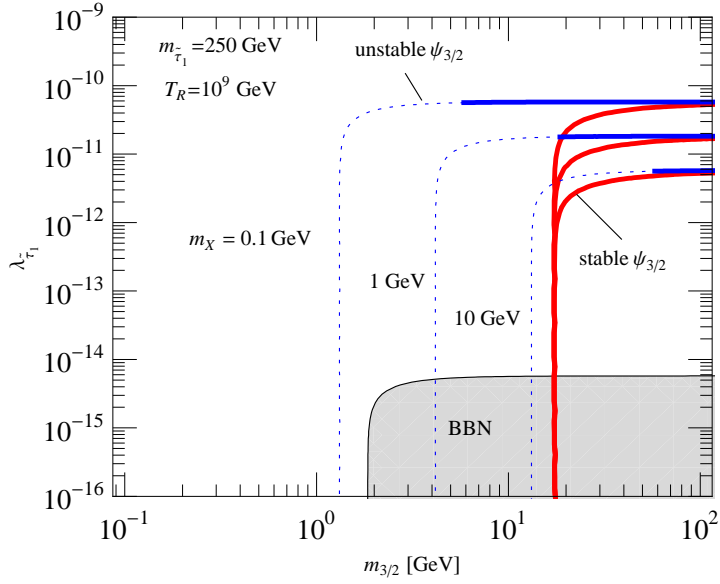}
  \end{center}
  \caption{Summary of constraints on stau LOSP scenario with \textit{hidden
  fermion LSP} and gravitino NLSP, similar to Fig.~\ref{fig:summaryI}.  Here,
  the mass of the hidden fermion $X$ is assumed to lie in the range
  $m_X=0.1$--$10\GeV$, whereas the reheating temperature and the $\lstau$ mass
  are fixed as indicated. The \textit{red} and \textit{blue lines} show the
  constraints for the case of a stable or unstable gravitino, respectively.
  \textit{Dashed} parts of the lines are again excluded by constraints on
  mixed warm/cold dark matter.}
  \label{fig:summaryII}
\end{figure}
In some scenarios the hidden fermion could be the LSP. If this is the case,
the gravitino can decay into it, if kinematically allowed, yielding a scenario
which is qualitatively different to the one studied in the previous
subsection. The total dark matter abundance is given in this case by:
\begin{align}
  \Omega_\text{dm}=\Omega_\hf^\text{th} +
  \Omega_\hf^{\lstau} +
  \underbrace{\frac{m_X}{m_{3/2}}\left(\Omega_{3/2}^\text{th} +
  \Omega_{3/2}^{\lstau}\right)}_{=\Omega_\text{WDM}}\;,
  \label{}
\end{align}
where we assumed that hidden-sector by-products of the gravitino decaying into
$X$ are massless.\footnote{Note that here again the component
$\Omega_X^\lstau$ has a large free-streaming length $\lambda_{FS}\gtrsim1\Mpc$
in some cases, depending on the coupling $\lambda_\lstau$ and the mass
$m_\lstau$. However, due to the small stau (and later the neutralino) yield,
the impact of this component is always negligible in our plots.} In this case,
the thermal production of gravitinos itself produces ultimately WDM, yielding
the constraint
\begin{align}
  \Omega_{3/2}^\text{th}h^2\lesssim \frac{m_{3/2}}{m_\hf}\;f\;0.11\;,
  \label{}
\end{align}
where we again assumed that $\Omega_{3/2}^{\lstau}\simeq0$. It is apparent
from this equation that in scenarios with unstable gravitinos and high
reheating temperatures, small masses $m_\hf\lesssim f\ m_{3/2}$ are favored, to
avoid strengthening the bounds on thermal gravitino production. Furthermore,
the thermal or non-thermal production of $\hf$ becomes essential, since it
must yield the dominant part of the dark matter abundance according to
$\Omega_\hf^\text{th}+\Omega_\hf^{\lstau} \simeq h^{-2}\ 0.11$, which implies a
coupling to the hidden fermion like $\lambda_{\lstau}\approx
10^{-12}(m_\lstau/m_X)^{1/2}$, as long as $\Omega_X^{\lstau}$ is negligible.

This situation is illustrated by the blue lines in Fig.~\ref{fig:summaryII},
for fixed reheating temperature $T_R=10^9\GeV$ and for different masses of the
hidden fermion $m_\hf$. As apparent from this plot, for being in agreement
with the bounds on mixed warm/cold dark matter the gravitino mass has to be
considerably higher than $m_X$ in each of the shown cases, \textit{e.g.}~for
$m_X=1\GeV$ the gravitino mass should exceed $20\GeV$. Typical couplings where
the mechanism works lie in the range $\lambda_\lstau\sim10^{-12}$--$10^{-10}$.

\subsection{Neutralino LOSP}
We will now briefly discuss the case of neutralino LOSP. For definiteness, we
consider an effective interaction Lagrangian between the hidden fermion $\hf$
and the lightest neutralino $\neut$ given by
\be
  -{\cal L} = g_{h}\, \bar \hf\,\neut\,h^0
  + g_{Z}\, \bar \hf\,\gamma_\mu \neut\,Z^\mu
  + g_{\gamma}\, \bar \hf\,\gamma_\mu \neut\, A^\mu 
  {+\text{h.c.}}
  \;.
  \label{eqn:Lagrangian2}
\ee
The couplings $g_i$ include gauge couplings, weak mixing angles etc.  The
neutralino can decay either $\neut \rightarrow (Z^0,\gamma,h^0)\psi_{3/2}$ or
$\neut \rightarrow (Z^0,\gamma,h^0)\hf$.

The electromagnetic and hadronic energy $\epsilon_{vis} = (m_{\neut}^2 -
m_{\psi_{3/2}/\hf}^2 + m_{Z/h}^2)/(2m_{\neut})$ released during these decays
can induce photo- and hadrodissociation of $^4$He during BBN, and can lead to
inter-conversions of protons and neutrons~\cite{Kawasaki:2004qu,
Jedamzik:2006xz}. Without hidden fermion $X$ (corresponding to the limits $g_i
\rightarrow 0$), this leads to stringent constraints on the gravitino mass and
consequently also on the reheating temperature \cite{Feng:2004mt}.  However,
if the couplings $g_i$ are large enough, the neutralino decays into hidden
fermions already before BBN, allowing a high reheating temperature $T_R \sim
10^9\GeV$ as required by leptogenesis.

As an example we will consider the decay $\neut \rightarrow Z^0 \hf$, where
the decay width is given by
\bea
  \Gamma_{\neut\to  Z^0 \hf} & = &
     {g_Z^2\over 32\pi} m_{\neut}\left[1+{m_\hf^2\over m_{\neut}^2}-2
     {m_Z^2\over m_{\neut}^2}+
     {m_{\neut}^2\over m_Z^2}\left(1-{m_\hf^2\over m_{\neut}^2}\right)^2-6
     {m_\hf\over  m_{\neut}}\right] \nonumber\\
  && \times\sqrt{\left[1-{(m_\hf+m_Z)^2\over m_{\neut}^2}\right]
     \left[1-{(m_\hf-m_Z)^2\over m_{\neut}^2}\right]}\,,
\eea
assuming that only $g_Z \not=0$. The $Z^0$ boson then decays into hadrons with
branching ratio $B_h\sim 0.7$.

For a reference neutralino yield of $\epsilon_{vis}Y_{\neut} \sim
100\GeV\times 10^{-12}$ \cite{ArkaniHamed:2006mb}, the most stringent
constraints come from overproduction of $^4$He due to interconversion
processes, as well as D production by hadrodissociation, leading to an upper
bound on the neutralino life-time in the range $\tau_{\neut} < \tau_{\rm max}
\sim 1 - 100\s$ \cite{Kawasaki:2004qu, Jedamzik:2006xz}.  In the parameter
region consistent with thermal leptogenesis, the decay into gravitinos is
negligible, and $\BR(\neut \rightarrow Z^0\hf)\simeq1$. This implies a lower
bound on the coupling, which reads in the limit $m_\neut\gg m_Z, m_X$:
\be
|g_Z| \gtrsim 
3\times 10^{-14} \left({m_{\neut}\over 200 \GeV}\right)^{-3/2}
\left(\frac{\tau_{\rm max}}{100\s}\right)^{-1/2} \;.
\ee

\begin{figure}[t]
  \begin{center}
    \includegraphics[width=0.45\linewidth]{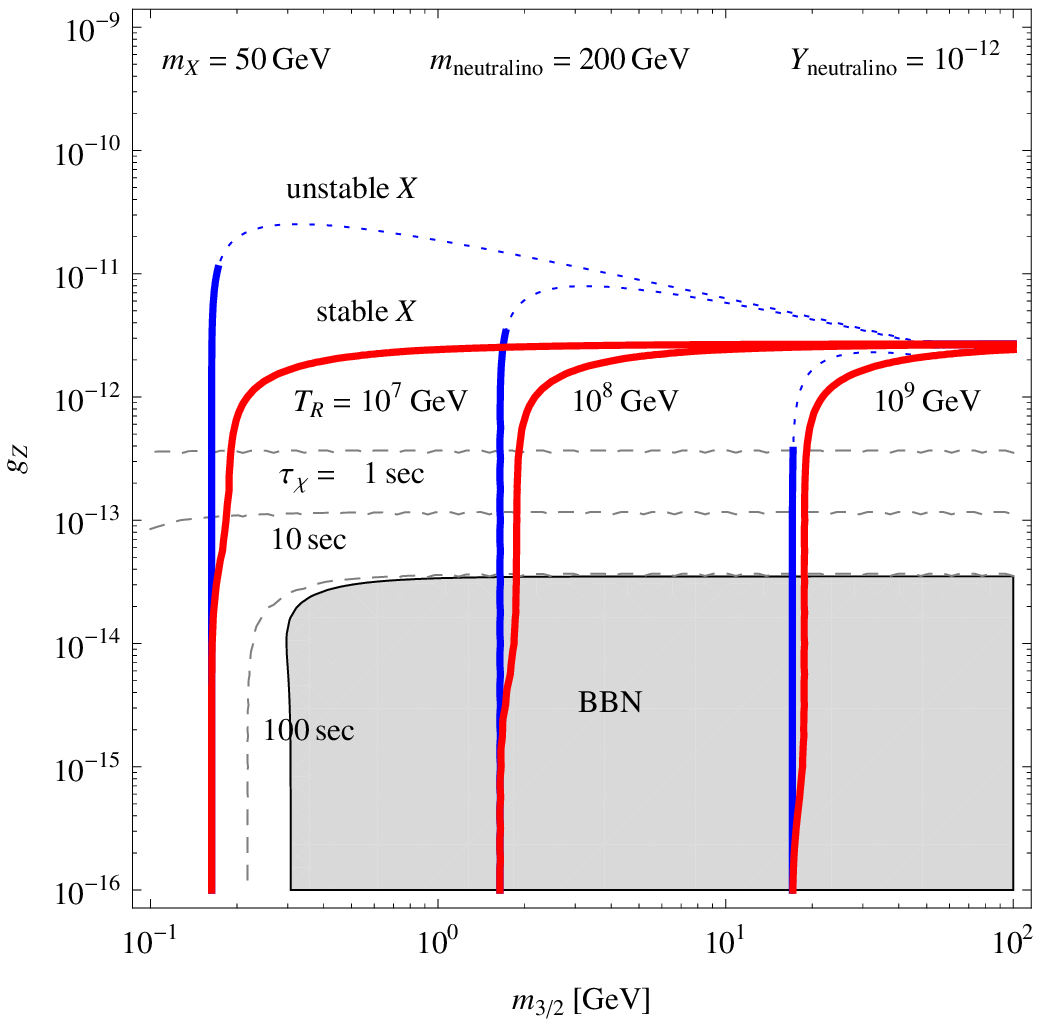}
    \includegraphics[width=0.45\linewidth]{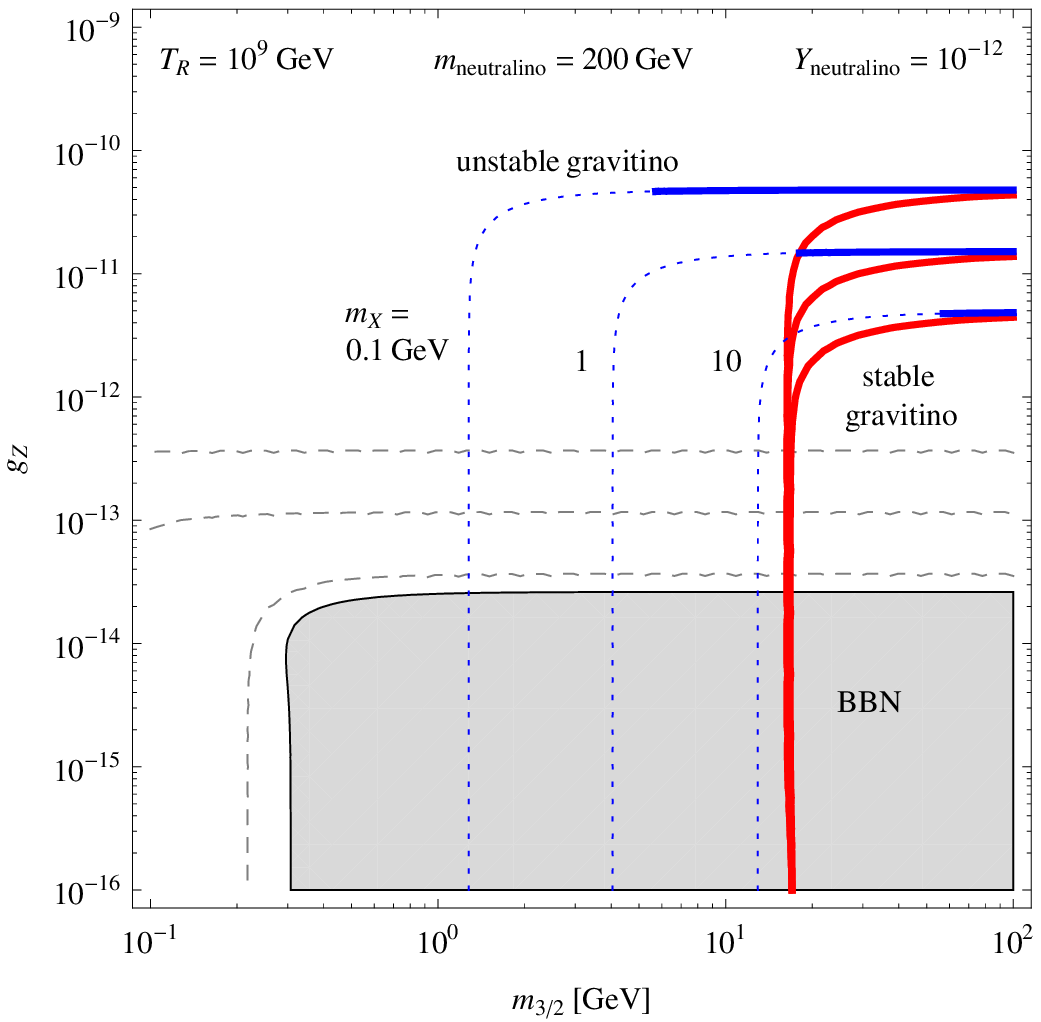}
  \end{center}
  \caption{ Summary of constraints on neutralino LOSP scenario with gravitino
  LSP (\textit{left panel}, as in Fig.~\ref{fig:summaryI}), and hidden fermion
  LSP (\textit{right panel}, as in Fig.~\ref{fig:summaryII}), as function of
  the $\neut \hf Z^0$ coupling $g_Z$ and gravitino mass $m_{3/2}$, and for
  $Y_\neut=10^{-12}$. The constraints from BBN obtained in
  Ref.~\cite{Jedamzik:2006xz} (\textit{shaded region}) exclude neutralino
  life-times $\tau_{\neut}\gtrsim100\s$. However, Ref.~\cite{Kawasaki:2004qu}
  presents somewhat stronger bounds, of the order $1-10\s$ (\textit{grey
  dashed lines}).}
  \label{fig:summaryIII}
\end{figure}

The constraints from overclosure and from free-streaming are very similar to
the case of stau LOSP discussed before, and we do not repeat them here. The
corresponding constraints for $Y_\neut=10^{-12}$ are summarized in
Fig.~\ref{fig:summaryIII}. In this figure, we also took the decay modes $\neut
\rightarrow (Z^0,\gamma)\psi_{3/2}$ into account using the rates given in
\cite{Feng:2004mt}, assuming a bino-like neutralino.\footnote{The BBN bounds
were obtained by interpolating between the bounds for the different hadronic
branching ratios given in \cite{Jedamzik:2006xz}.} Note that for both cases,
gravitino and hidden fermion LSPs, one can find parameters compatible with
$T_R \sim 10^9\GeV$, and typical mixing parameters lie in the range
$g_Z\sim10^{-13}$--$10^{-11}$.

The bounds shown in Fig.~\ref{fig:summaryIII} are relying on the relatively
small adopted neutralino yield, and they can change qualitatively when the
yield is much larger. For example, in the case where the hidden fermion is the
NLSP and unstable (blue lines in the left panel of Fig.~\ref{fig:summaryIII}),
the warm dark matter component produced by the decay chain $\neut \rightarrow
\hf \rightarrow \psi_{3/2}$ implies an upper limit on the neutralino yield,
\begin{equation}
  Y_{\neut} \lesssim 
  10^{-12} \, \frac{20\GeV}{m_{3/2}} \,
  \frac{f}{0.05} \;,
\end{equation}
which follows from the bounds on mixed warm/cold dark matter. On the other
hand, in the cases where the gravitino is the NLSP, one can find viable
scenarios even for much higher values of the neutralino yield, which however
requires that the hidden fermion is very light.

\section{Examples}
After having discussed the proposed mechanism as well as astrophysical and
cosmological bounds in the last section in general, we will now present two
concrete scenarios with hidden fermions from a vector and from a chiral
supermultiplet, respectively.

\subsection{Hidden gauginos of an unbroken \boldmath{$U(1)_X$}}
Let us consider the case where the hidden fermion arises from the gaugino
component of a \textit{vector superfield} of an unbroken hidden $U(1)_X$
symmetry, which mixes with the $U(1)_Y$ of hypercharge via a small kinetic
mixing $\chi\ll1$ (for details see also Ref.~\cite{Ibarra:2008kn}). This
scenario is an example for the case where the superpartner of the hidden
fermion $X$, here the $U(1)_X$ vector boson, remains exactly massless. For
simplicity, we will assume that all matter charged under $U(1)_X$ is
vector-like and heavy enough to be cosmologically irrelevant.

When SUSY is exact, the canonical normalization of the kinetic terms of the
vector superfields produces an unobservable shift of the hypercharge gauge
coupling, while the hidden $U(1)_X$ gauge boson and gaugino completely
decouple from the observable sector~\cite{Holdom:1985ag}.  However, in
presence of SUSY breaking effects, the decoupling of the gaugino is not
complete any more~\cite{Ibarra:2008kn}. In the component formalism, the
relevant part of the Lagrangian, including the supersymmetry breaking soft
masses, reads
\begin{eqnarray}
  \label{eqn:Lagrangian}
  \mathcal{L}_{gauge}&
  =&-\frac{1}{4} \ppp{\hat{X}_{\mu\nu}&\hat{B}_{\mu\nu}} \mathcal{K} 
  \ppp{\hat{X}^{\mu\nu}\\ \hat{B}^{\mu\nu}}
  -i\ppp{ \hat{\lambda}_X & \hat{\lambda}_B}\mathcal{K}
  \sigma^\mu\partial_\mu\ppp{\hat{\lambda}_{X}^\dagger\\
  \hat{\lambda}_{B}^\dagger}
  \\\nonumber&& 
  +\frac{1}{2}\ppp{\hat{D}_X &
  \hat{D}_B}\mathcal{K}\ppp{\hat{D}_{X}\\\hat{D}_{B}}
  -\left[\frac{1}{2}\ppp{\hat{\lambda}_X &
  \hat{\lambda}_B}\hat{\mathcal{M}}
  \ppp{\hat{\lambda}_{X}\\\hat{\lambda}_{B}} + \text{h.c.}\right]\;,
\end{eqnarray}
where $\mathcal{K}$ and $\hat{\mathcal{M}}$ denote, respectively, the kinetic
and mass mixing matrices 
\begin{eqnarray}
  \mathcal{K}=\ppp{1&\chi\\\chi&1}\hspace{0.5cm}\text{and}\hspace{0.5cm}
  \hat{\mathcal{M}}=\ppp{\hat{M}_X&\delta\hat{M}\\\delta\hat{M}&\hat{M}_B}\;,
  \label{eqn:MixingMatrix}
\end{eqnarray}
and $\hat\lambda_{X/B}$ and $\hat D_{X/B}$ are the gauginos and D-terms
corresponding to the gauge fields $\hat X_{\mu\nu}$ and $\hat B_{\mu\nu}$,
respectively. Note that the generation of mass mixing in general depends on
details of the underlying theory.

In the basis where the kinetic terms are canonical, which can be achieved by
the redefinition (which also holds for the corresponding $D$-terms and the
vector bosons)
\begin{eqnarray}
  \ppp{\hat{\lambda}_X\\\hat{\lambda}_B}=
  \ppp{1&-\frac{\chi}{\sqrt{1-\chi^2}}\\0&\frac{1}{\sqrt{1-\chi^2}}}
  \ppp{\tilde\lambda_X \\\tilde \lambda_B}\;,
  \label{eqn:GL2Transformation}
\end{eqnarray}
the hidden gaugino and the four MSSM neutralinos mix, and the corresponding
extended $(5\times5)$ neutralino mass matrix reads, to lowest order in $\chi$, 
\begin{eqnarray}
  \mathcal{M}_\text{N}=\ppp{ M_X & \delta M & 0 & 0 & 0 \\ 
  \delta M & M_B & 0 & -M_Z c_\beta s_W & M_Z s_\beta s_W \\
  0 & 0 & M_W & M_Z c_\beta c_W & -M_Z s_\beta c_W \\
  0 & -M_Z c_\beta s_W & M_Z c_\beta c_W & 0 & -\mu \\
  0 & M_Z s_\beta s_W & -M_Z s_\beta c_W & -\mu & 0 }\;,
  \label{eqn:NeutralinoMassMatrix}
\end{eqnarray}
where $\delta M \simeq \delta\hat{M} -\chi \hat{M}_X$, $M_X\simeq\hat{M}_X$
and $M_B\simeq\hat{M}_B$.  Here, $\mu$ denotes the MSSM $\mu$-term, $M_Z$ the
mass of the $Z^0$ gauge boson, $s_W$ the sine of the Weinberg angle and
$s_\beta$ is related to the ratio of the two Higgs VEVs. 

Upon diagonalization of the previous mass matrix we find a mass eigenstate
$X=\tilde \lambda_X+\Theta \tilde \lambda_B$ with mass $m_X\simeq M_X$, namely
it is mostly hidden fermion with a small bino admixture given by
\begin{eqnarray}
  \Theta \simeq
  \frac{\delta M}{M_B-M_X}\;,
  \label{eqn:ChiEffective}
\end{eqnarray}
from which follows that typically $\Theta\sim\mathcal{O}(\chi)$. Thus, even
though the canonically normalized fields $\tilde X_{\mu\nu}$ and $\tilde D_X$
completely decouple from the observable sector, the mass eigenstate $X$,
couples to the observable sector (\fex~the right-handed stau $\tilde\tau_R$)
through the tiny bino component
\begin{equation}
  -{\cal L}\supset \sqrt{2} g' Y_{\tilde\tau_R}( \bar{\tilde \lambda}_B
  P_R\tau) \stau_R \rightarrow \sqrt{2} g' Y_{\tilde\tau_R} \Theta (\bar  X
  P_R\tau) \stau_R \;.
\end{equation}

The presence of an additional $U(1)_X$ gauge group opens the stau decay
channel $\lstau\to\tau X$ (we assume that $\lstau\simeq \tilde\tau_R$), with
lifetime
\begin{align}
  \Gamma_{\lstau\to\tau X}=
  \frac{g'^2}{8\pi}\Theta^2 Y_{\tilde\tau_R}^2 m_{\lstau}
  \left( 1-\frac{m_X^2}{m_{\lstau}^2} \right)^2.
  \label{eqn:StauLifetime}
\end{align}
In contrast to the scenario discussed in Sec.~2, where only the minimal
effects of the coupling~\eqref{eqn:Lagrangian1} where taken into account, many
additional channels for thermal production are now open, since the $U(1)_X$
inherits all couplings of the hypercharge bino. The thermal production of
hidden gauginos from $2\to2$ scattering processes was calculated
in~Ref.~\cite{Ibarra:2008kn} and yielded an upper bound on the mixing angle
$\Theta$ that is given by (provided that $X$ is stable)
\begin{align}
  \Theta \lesssim 3\times 10^{-12}\sqrt{\frac{m_{\tilde q}}{m_X}}\;,
  \label{eqn:ThetaBound1}
\end{align}
where $m_{\tilde q}$ denotes the squark masses.  In this calculation the
dominant scattering processes involving one QCD and one hypercharge vertex
were taken into account. Additional contributions from $1\to2$ processes were
discussed in Ref.~\cite{Hall:2009bx}, yielding a relic density according to
Eq.~\eqref{eqn:Xproduction1}, when identifying
$\lambda_{\lstau}^2=g'^2\,Y^2_{\widetilde\tau_R}\,\Theta^2$ and
\begin{align}
  \delta = \sum_{\tilde f\neq \lstau} 
  \frac{Y_{\tilde f}^2}{Y_{\tilde\tau_R}^2}\frac{m_{\lstau}}{m_{\tilde f}}
  \left( \frac{1- m_X^2/m_{\tilde f}^2}{1-m_X^2/m_{\lstau}^2} \right)^2\;,
  \label{eqn:DefDelta}
\end{align}
where $\tilde f$ runs over all sfermions of the MSSM. For a typical SUSY mass
spectrum, one obtains $\delta\sim\mathcal{O}(3$--$10)$. For stable $X$, this
yields an upper bound of
\begin{align}
  \Theta \lesssim 10^{-12}\sqrt{\frac{m_{\lstau}}{m_X}}\;.
  \label{}
\end{align}
Note that both contributions from $1\to2$ and $2\to2$ processes are roughly of
the same order of magnitude and not much stronger than the production coming
only from a term like Eq.~\eqref{eqn:Lagrangian1} alone.

\begin{figure}[t]
  \begin{center}
    \includegraphics[width=0.45\linewidth]{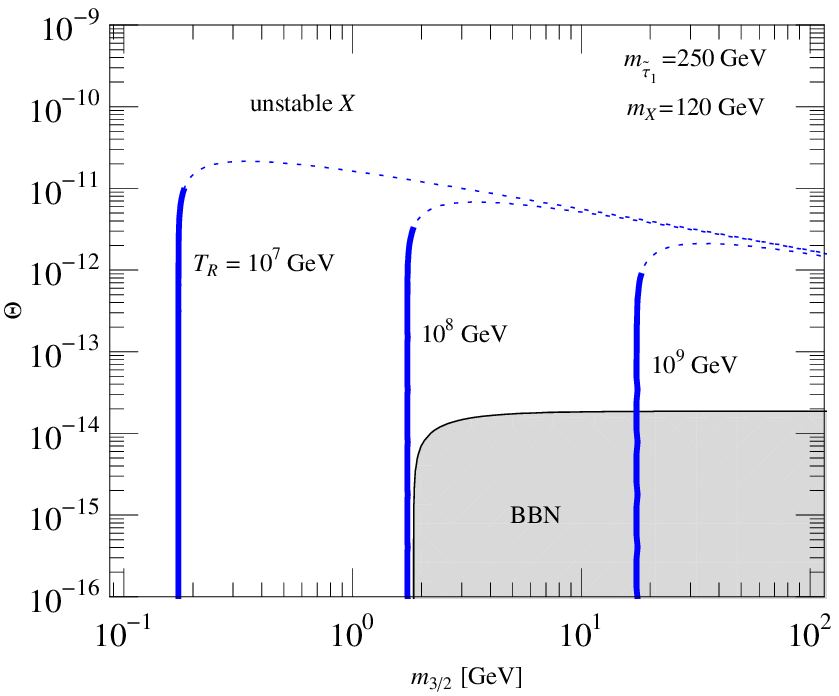}
    \includegraphics[width=0.45\linewidth]{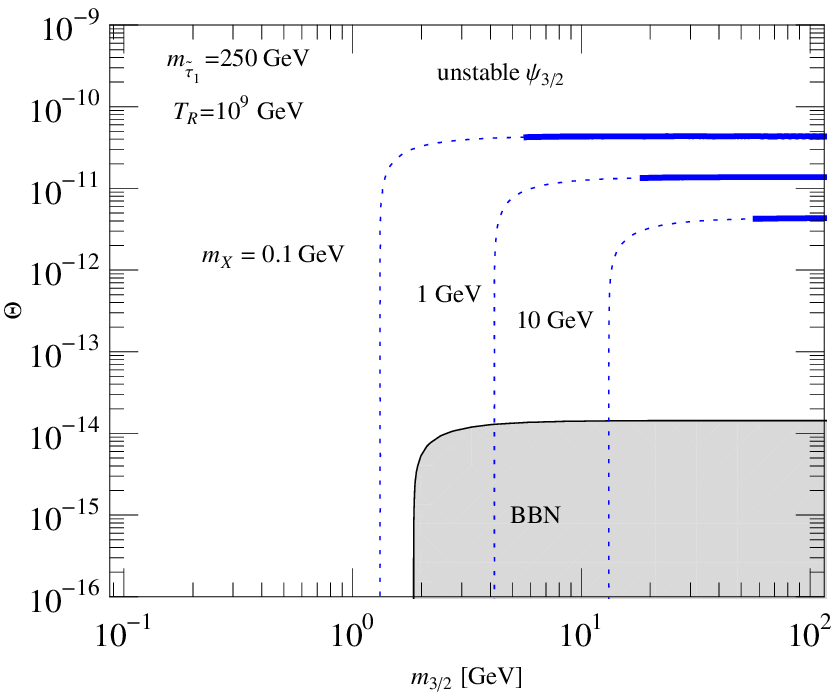}
  \end{center}
  \caption{Constraints on the parameter space of the hidden $U(1)_X$ gaugino
  scenario with a $\lstau$ LOSP, for the case of gravitino LSPs (\textit{left
  panel}, \textit{cf.}~Fig.~\ref{fig:summaryI}) and for hidden gaugino LSPs
  (\textit{right panel}, \textit{cf.}~Fig.~\ref{fig:summaryII}), as function
  of the mixing parameter $\Theta$ and the gravitino mass $m_{3/2}$, and for
  $\lstau$ mass fixed as indicated. In the \textit{left panel} the reheating
  temperature varies and whereas in the \textit{right panel} the hidden
  gaugino mass $m_X$ varies. In both cases the NLSP is unstable and decays
  into the LSP, since the hidden vector boson remains massless.}
  \label{fig:hg}
\end{figure}

Bounds on the mixing parameter $\Theta$ are illustrated in Fig.~\ref{fig:hg}.
From there it is clear that a kinetically mixed hidden $U(1)_X$ gauge group
with mixing parameters $10^{-13}\lesssim \Theta \lesssim 10^{-10}$ satisfies
all constraints. Apart from the somewhat stronger thermal production, the
situation is similar to what is shown in Fig.~\ref{fig:summaryI}
and~\ref{fig:summaryII}.

The required mixing parameter $\Theta\sim\chi$ lies in the broad range of
values that can be accommodated in string motivated $U(1)_X$ extensions:
Without additional symmetries, the kinetic mixing $\chi$ is generically
generated on one-loop level by integrating out chiral superfields charged
under both, visible and hidden sector. In this case it acquires values
typically around $\chi\sim 10^{-4}$--$10^{-2}$, corresponding to one-loop
suppression~\cite{Holdom:1985ag, delAguila:1988jz}. However, in
compactifications of heterotic and type~II strings, much smaller mixings are
possible~\cite{Dienes:1996zr, Abel:2008ai, Goodsell:2009xc, Arvanitaki:2009hb,
Goodsell:2010ie}. For example, a lower bound around $\chi \gtrsim 10^{-16}$
was argued to hold in cases of gauge mediated supersymmetry breaking in
heterotic string models \cite{Dienes:1996zr}, whereas in type-II models with
warped extra dimensions the kinetic mixing parameter could be parametrically
even smaller~\cite{Abel:2008ai}. Hence, scenarios with hidden unbroken
$U(1)_X$ gauge groups provide simple and natural scenarios where the tension
between thermal leptogenesis and gravitino dark matter is
solved~\cite{Ibarra:2008kn}.

\subsection{A model for small couplings from non-renormalizable operators}
\label{Model}
Let us now consider the case where the hidden fermion $\hf$ is part of a
\textit{chiral supermultiplet}, singlet under the Standard Model gauge group.
Then, the gauge symmetry allows the superpotential term
\begin{equation}
  W \supset \frac{\alpha}{M} L H_d e^c_{R} \hf\;.
\label{wanted-W}
\end{equation}
This term leads, after the electroweak symmetry breaking, to a Lagrangian term
of the form Eq.~\eqref{eqn:Lagrangian1}, with a Yukawa coupling which reads
\begin{equation}
  \lambda_\lstau \sim \frac{\alpha_{33} \langle H_d\rangle}{M} \sim
  10^{-14} \alpha_{33}\cos\beta
  \left(\frac{M}{10^{16}\GeV}\right)^{-1}\;.
  \label{natural-range}
\end{equation}
It is remarkable that this value for the
coupling lies in the allowed region of Figs.~\ref{fig:summaryI} and
\ref{fig:summaryII} if the non-renormalizable operator is suppressed by masses
close to the Grand Unification scale.

The superpotential Eq.(\ref{wanted-W}) can naturally arise from the decoupling
of heavy particles with a mass of ${\cal O}(M)$.  Let us consider an extension
of the MSSM with three additional chiral superfields: $\hf~(1,1,0)$,
$H'_u~(1,2,1)$ and $H'_d~(1,2,-1)$, where in parenthesis we indicate the
quantum numbers under $SU(3)_c\times SU(2)_L\times U(1)_Y$.  Note that $H'_u$
and $H'_d$ have identical gauge quantum numbers as the MSSM Higgs doublets
$H_u$ and $H_d$, respectively. To avoid unwanted terms, we will further impose
the following Peccei-Quinn transformation on the superfields:
\begin{eqnarray}
  (Q,U^c_R,D^c_R,L,E^c_R,N^c_R) & \rightarrow & e^{i\alpha}
  (Q,U^c_R,D^c_R,L,E^c_R,N^c_R)\,, \\
  (H_u,H_d)             
  & \rightarrow & e^{-2i\alpha}(H_u,H_d) \,,\\
  H_u' & \rightarrow & e^{2i\alpha}H_u' \,, \\
  H_d' & \rightarrow & e^{-2i\alpha}H_d' \,, \\
  \hf & \rightarrow & \hf \,.
\end{eqnarray}
This symmetry forbids the bilinear term $H_u H_d$ in the superpotential as
well as a Majorana mass term for the right-handed neutrinos. The bilinear
$\mu$ term could be generated via the Giudice-Masiero
mechanism~\cite{Giudice:1988yz}. On the other hand, in order to generate the
right-handed neutrino masses, we will further introduce a Standard Model
singlet, $\Phi$, which transforms under the Peccei-Quinn symmetry as
$\Phi\rightarrow e^{-2 i \alpha}\Phi$.  Then, the term $\Phi N_R^c N_R^c$ is
allowed in the superpotential and leads to right-handed Majorana masses if the
scalar component of $\Phi$ acquires a vacuum expectation value, $\langle \Phi
\rangle \sim M_R$.  Note that even with the presence of the new field it is
not possible to generate a bilinear term $H_u H_d$ in the
superpotential.\footnote{In the presence of the field $\Phi$ new,
non-renormalizable, terms appear in the superpotential suppressed by a large
mass scale, $M_*$, such as $QQQL\Phi^2/M^3_*$.  After the breaking of the
Peccei-Quinn symmetry this term leads to a superpotential term $(M_R/M_*)^2
QQQL/M_*$ which induces proton decay. However, the small factor $(M_R/M_*)^2$
and the plausibly small coefficients of the dimension-7 operator for the first
generation can yield a proton lifetime in agreement with the stringent
experimental bounds.}

Then, the renormalizable superpotential of the model reads
\begin{eqnarray}
  W & = & L (\lambda_E \cdot H_D) E^c_R + Q (\lambda_D \cdot H_D) D^c_R +
  \lambda_uQH_uU^c_R+ \lambda_\nu L H_u N^c_R + \lambda_{\rm M} \Phi N^c_R
  N^c_R \nonumber \\ &   & {} + (\mu_D \cdot H_D ) H_u' + (\lambda_\hf \cdot
  H_D ) H_u' \hf + M_\hf \hf\hf + \kappa_\hf \hf^3 + C_\hf \hf + \dots \,,
\end{eqnarray}
where the ellipsis indicates additional terms in the superpotential which lead
to the breaking of supersymmetry and the breaking of the Peccei-Quinn
symmetry. Here, we used a compact notation where $H_D=(H_d,H_d')$, $\lambda_E
\cdot H_D = \lambda_e H_d + \lambda_e' H_d'$, $\lambda_X \cdot H_D = \lambda_x
H_d + \lambda_x' H_d'$ etc.  The linear term in $\hf$ can be eliminated by a
shift, $\hf \rightarrow \hf - C_\hf/(2M_\hf)$.  Upon redefining the
parameters, this amounts to setting $C_\hf=0$. In general $\mu_{Di}\sim{\cal
O}(M_{\rm GUT})$, although we will choose to work in the basis in the
$(H_d,H_d')$-space such that $\mu_D=(0,M)$, being $M\sim {\cal O}(M_{\rm
GUT})$. Written in components, we thus have
\begin{eqnarray}
  W & = & W_{\rm MSSM}^{\rm Yukawa}+\lambda_\nu L H_u N^c_R  +
  \lambda_{\rm M} \Phi N^c_R N^c_R  \nonumber \\
  &   & {} +\lambda_e' L  H_d' E_R^c + \lambda_d' Q H_d' D_R^c
  + M H_d' H_u' + (\lambda_x H_d + \lambda_x' H_d' ) H_u' \hf  \nonumber \\
  &   & {} + M_\hf \hf\hf + \kappa_\hf \hf^3 + \dots  \,.
  \label{eqn:PQlagMod}
\end{eqnarray}

The breaking of the Peccei-Quinn symmetry at intermediate scales leads to
Majorana masses for the right-handed neutrinos, $M_{\rm M}$.  Then, the
decoupling of the heavy fields $H_u'$, $H_d'$, $N_R^c$ finally leads to the
following effective superpotential
\begin{align}
  W_{\rm eff}=&W_{\rm MSSM}^{\rm Yukawa}- (\lambda_\nu M^{-1}_M \lambda_\nu^T)
  (L H_d)(LH_d)\nonumber\\
  &- \frac{\lambda_e' \lambda_x}{M} L H_d E^c_{R} \hf -
  \frac{\lambda_d' \lambda_x}{M} Q H_d D^c_{R} \hf + M_\hf \hf\hf +
  \kappa_\hf \hf^3\;,
  \label{eqn:lag2}
\end{align}
which contains the term Eq.~(\ref{wanted-W}). Note that this model is free
from gauge anomalies and preserves the successful MSSM gauge coupling
unification. 

In this model, there are additional contributions to the thermal production of
the hidden fermion $X$ from $2\to2$ scattering processes, such as $\stau\
h^0\rightarrow \hf\ \tau$ and $\stau\ \tau\rightarrow \hf\ h^0$. These
processes stem from the higher-dimensional operators (HDO) in
Eq.~\eqref{eqn:lag2}, and the amplitude for these processes increases with
energy squared, $|\mathcal{M}|^2_{HDO} \sim  s / M^2$, where $s$ denotes the
square of the center-of-mass energy.  Therefore these production channels can
be very efficient at the high temperatures required by thermal leptogenesis.
We will now shortly discuss the impact of these production channels on the
overproduction constraints.

The sum of the matrix elements for all possible scatterings producing an $\hf$
or an $\tilde \hf$, respectively, are
\begin{equation}
  \sum_{channels} |\mathcal{M}|^2_{12\rightarrow 3\hf} = \sum_{channels}
  |\mathcal{M}|^2_{12\rightarrow 3\tilde \hf} = 12s \times
  \frac{\lambda_x^2}{M^2} \sum_{i,j} \left(  (\lambda_e')_{ij}^2 + 3
  (\lambda_d')_{ij}^2\right) \;,
\end{equation}
where we also summed over the initial and final-state spins. The effective
$\hf$-stau-tau coupling can be identified with
\begin{equation}
  \lambda_\lstau^2 \equiv \frac{\lambda_x^2v_d^2}{M^2} \sum_i
  (\lambda_e')_{i\tau}^2 \equiv \frac{\alpha^2v_d^2}{M^2} \;,
\end{equation}
where $v_d=v\cdot\cos\beta\simeq\cos\beta\ 175\GeV$. Assuming for simplicity
that all other Yukawa couplings are zero (which will give the minimal
contribution to the UV thermal production), the hidden fermion yield
reads~\cite{Hall:2009bx} 
\begin{eqnarray}
  \label{eqn:HFyield}
  Y_{HDO} & \simeq & 3\times 12\times \frac{0.4 T_R \alpha^2
  \sqrt{8\pi}M_P}{\pi^7M^2g_*^{3/2}} \\ & =      & 1.7 \times 10^{-10}
  \alpha^2 \left(\frac{T_R}{10^{9}\GeV}\right)
  \left(\frac{10^{16}\GeV}{M}\right)^2 \left(\frac{915/4}{g_*}\right)^{3/2}\;,
  \nonumber
\end{eqnarray}
where $M_P\simeq 2.4\times 10^{18}\GeV$ denotes the reduced Planck mass. In
this expression the prefactor $3$ takes into account that the hidden fermions
are produced in two ways: the scattering  $12\rightarrow 3 X$, which produces
just one hidden fermion, and the scattering $12\rightarrow 3 \tilde X$, which
produces {\it two} hidden fermions, due to the fast decay  $\tilde X\to XX$
(unless $\kappa_X\ll10^{-12}$). Assuming that $\hf$ is stable on cosmological
time-scales, this leads to an abundance
\begin{equation}
  \Omega_\hf^{HDO}h^2 \simeq 0.045\alpha^2
  \left(\frac{m_\hf}{1\GeV}\right) \left(\frac{T_R}{10^{9}\GeV}\right)
  \left(\frac{10^{16}\GeV}{M}\right)^2 \left(\frac{915/4}{g_*}\right)^{3/2}\;.
\end{equation}
For comparison with the freeze-in abundance~\eqref{eqn:Xproduction1}, it is
instructive to express this in terms of the effective coupling parameter
$\lambda_\lstau$. Then, with $g_\ast=915/4$,
\begin{equation}
  \Omega_\hf^{HDO}h^2 \simeq 0.11 
  \cos^{-2}\beta
  \left(\frac{m_\hf}{1\GeV}\right)
  \left(\frac{T_R}{10^{9}\GeV}\right) \left(\frac{\lambda_\lstau}{2.7\times
  10^{-14}}\right)^2 \;.
\end{equation}
From requiring $\Omega_\hf^{HDO}h^2 < 0.11$ one thus obtains a bound
\begin{equation}
  |\lambda_\lstau| \lesssim 2.7\times 10^{-14}
  \cos\beta
  \left(\frac{1\GeV}{m_\hf}\right)^\frac12
  \left(\frac{10^{9}\GeV}{T_R}\right)^\frac12 \;,
\end{equation}
which depends on the reheating temperature $T_R$ and the mass of the hidden
fermion $m_X$.

\begin{figure}[t]
  \begin{center}
    \includegraphics[width=0.6\linewidth]{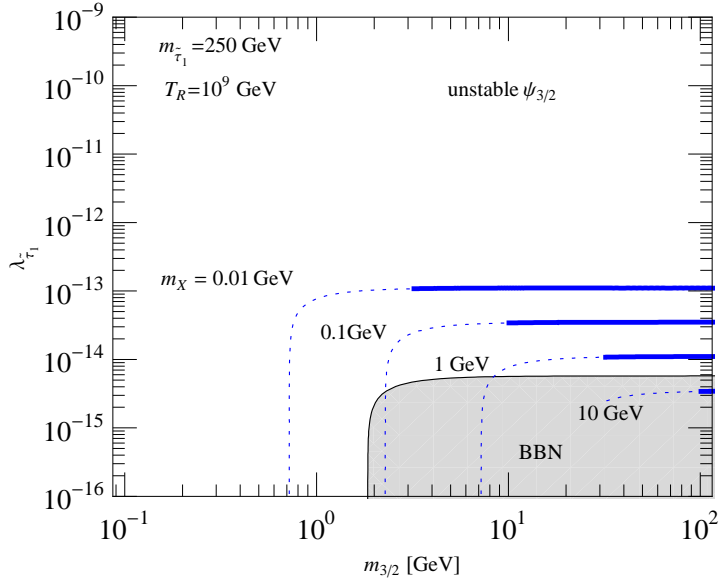}
  \end{center}
  \caption{Similar to Fig.~\ref{fig:summaryII}, but for the model described in
  Section.~\ref{Model}, taking into account also the minimal ultraviolet
  contributions to the $\hf$ production, and assuming $\tan\beta=2$. We show
  constraints for the case where the gravitino is the NLSP and decays into
  $\hf$ (in contrast to the previous figures the gravitino decays here into
  three hidden fermions, $\psi_{3/2}\rightarrow XXX$). The reheating
  temperature and the $\lstau$ mass are fixed to $10^9\GeV$ and $250\GeV$,
  respectively, whereas the hidden fermion $X$ mass varies in the range
  $m_X=0.01$--$10\GeV$. Note that a typical value for $\lambda_\lstau$ in our
  model is $\sim 10^{-14}$--$10^{-13}$ when the non-renormalizable operator is
  suppressed by masses close to the GUT scale, {\it cf.}
  Eq.(\ref{natural-range}).}
  \label{fig:ModelI}
\end{figure}

Bounds on the above model, for the case where the hidden fermion mass is very
small, are summarized in Fig.~\ref{fig:ModelI} (note that gravitino LSPs
together with reheating temperatures $T_R\sim10^9\GeV$ are excluded in the
present model due to overproduction of the hidden fermion NLSPs). We show
results for unstable gravitinos only, since even a very small $\kappa_X$ makes
the gravitino unstable at cosmological timescales, due to the fast decay
$\psi_{3/2}\to XXX$. When comparing Fig.~\ref{fig:ModelI} with
Fig.~\ref{fig:summaryII}, it is apparent that the contribution to the relic
abundance of hidden fermions coming from the HDO,
\textit{cf.}~Eq.~\eqref{eqn:HFyield}, reduces the allowed parameter space for
the coupling $\lambda_\lstau$ considerably. These ultraviolet contributions
depend linearly on the reheating temperature and dominate the infrared
contributions coming from the renormalizable operator. They can potentially
reintroduce the gravitino problem, \textit{cf.}~Eq.~\eqref{eqn:OmegaGravTP},
if the mass of the hidden fermion is too large. However, for small hidden
fermion masses $m_X\lesssim1\GeV$ all constraints from overproduction, BBN and
structure formation can be simultaneously satisfied for couplings in the range
$\lambda_\lstau\sim10^{-14}$--$10^{-13}$. Most interestingly, these couplings
are preferred if the non-renormalizable operator is suppressed by masses close
to the GUT scale, see Eq.~\eqref{natural-range}.

\section{Experimental signatures}
In this section we will briefly discuss possible experimental signatures of
the scenario proposed above at colliders and in cosmic-ray observations.

In this scenario the coupling constant of the LOSP to the hidden fermion $X$
is typically smaller than $10^{-12}$, therefore the decay length is much
larger than the size of typical collider detectors. As a consequence, if the
LOSP is the lightest neutralino, the experimental signatures at colliders are
identical to the case of the MSSM with R-parity conserved, since the lightest
neutralino just escapes the detector.  On the other hand, if the LOSP is the
lightest stau, it propagates through the detector leaving a heavily ionizing
charged track~\cite{Drees:1990yw}. Furthermore, if the stau velocity is small
enough, it could get trapped in the detector and decay eventually, producing a
tau moving in a non-radial direction.  This signature, albeit very
spectacular, is not specific of this scenario but also arises in scenarios
with stau LOSP and gravitino or axino LSP~\cite{stopped-staus}.

A hint towards our proposed scenario arises from the measurement of the
coupling stau-tau-hidden fermion. More concretely, at colliders it will be
possible a determination of the stau mass and the stau lifetime, from which
the coupling could be determined through Eq.~(\ref{eq:rate-into-X}):
\begin{equation}
  |\lambda_{\lstau}|\simeq \sqrt{\frac{8\pi}{\tau_\lstau m_\lstau}}\;.
\end{equation}
If the coupling inferred from collider experiments lies in the range preferred
by cosmology, Eq.~(\ref{eq:lambda-upperbound}), and if the lifetime
$\tau_\lstau$ is in agreement with BBN bounds, this scenario would gain
strength.

Furthermore, if the small coupling of the stau to the tau and the hidden
fermion is due to the non-renormalizable operator Eq.(\ref{wanted-W}), then a
very interesting signature could be observed at colliders: after being stopped
in the detector, the stau could decay producing also a Higgs boson, with
branching ratio:
\begin{equation}
  {\rm BR}(\lstau\to\tau h X)  \simeq  \frac{\tan^2\beta\,\sin^2\alpha
  }{384\pi^2} \frac{m_\lstau^2}{v^2} f(m_h^2/m_\lstau^2)\;,
\end{equation}
where $f(r)=1 + 9r -9r^2 -r^3 + 6r(1+r)\ln (r)$, $v=175\GeV$
is the Higgs vacuum expectation value and $\alpha$
is the angle that diagonalizes the CP-even Higgs squared-mass matrix.
For typical values of the light Higgs mass and the stau mass, 
we obtain
\begin{equation}
  {\rm BR}(\lstau\to\tau h X) \simeq  0.006
  \left(\frac{m_\lstau}{250\GeV}\right)^2 \left(\frac{\tan\beta}{10}\right)^2
  \sin^2\alpha \;,
\end{equation}
where we have taken $m_h/m_{\lstau}=115/250$ in the argument of the function
$f$.  Therefore, if 1000 staus could be stopped, a few Higgs events could be
observed in the detector.

For comparison, in scenarios with gravitino LSP and stau NLSP,
the Higgs decay is strongly suppressed by the tau Yukawa coupling.
The branching ratio, taking Higgs-strahlung into account, reads
\begin{equation}
  {\rm BR}(\lstau\to\tau h \psi_{3/2})  \simeq  \frac{\tan^2\beta\sin^2\alpha
  }{64\pi^2} \frac{m_\tau^2}{v^2} g(m_h^2/m_{\lstau}^2) \;,
\end{equation}
where $g(r) = (1+8r+6r^2)\ln (r^{-1}) - \left( 43 + 80r - 108r^2 - 16r^3 +
r^4\right)/12$, and we assumed $m_{3/2} \ll m_\lstau$.
Again, inserting typical values of parameters,
\begin{equation}
  {\rm BR}(\lstau\to\tau h \psi_{3/2})\simeq  3.4\times 10^{-7}
  \left(\frac{\tan\beta}{10}\right)^2 \sin^2\alpha\;,
\end{equation}
which is suppressed by the tau mass, making the observation of Higgs events
from stopped staus more difficult.  Thus, the number of Higgs events in late
stau decays constitutes a way to discriminate between a scenario with
gravitino/axino LSP and a scenario where a hidden fermion couples to the stau
and the tau via the non-renormalizable operator Eq.~(\ref{wanted-W}).\bigskip

Another possible experimental signature in this scenario is 
the observation of a large number of lepton flavour violating stau-LOSP 
decays. In the model described in section \ref{Model} it is in general 
not possible to simultaneously diagonalize the usual MSSM coupling 
$\lambda_e L H_d E^c_R$ and the new coupling with the heavy Higgs
doublet $\lambda'_e L H'_d E^c_R$. As a consequence, the effective
coupling in the Lagrangian $(\lambda'_e)_{ij} \lambda_x \langle H_d \rangle/M
\bar L_i \tilde E^c_{Rj} X$ in general contains
sizeable off-diagonal entries, which induce {\it at tree level} the
lepton flavour violating decays
$\lstau\rightarrow \mu X$ and  $\lstau\rightarrow e X$ with branching ratio:
\begin{equation}
  \frac{{\rm BR}(\lstau\to\ell_i X)}{{\rm BR}(\lstau\to\tau X)}
  = \left|\frac{(\lambda'_e)_{i3}}{(\lambda'_e)_{33}}\right|^2\;.
\end{equation}
Future colliders are capable of detecting the electrons and muons
produced in the decay of long lived staus if 
$(\lambda'_e)_{i3}/(\lambda'_e)_{33}$ is larger than 
$\sim 3\times 10^{-2}~(9\times 10^{-3})$ provided  
$3\times 10^3~(3\times 10^4)$ 
staus can be collected~\cite{Hamaguchi:2004ne}. Therefore, this scenario
offers good prospects to detect flavour violation in stau decays at
future colliders. 

It is important to note that this scenario can be easily compatible
with the present experimental constraints on lepton flavour violation
in the charged lepton sector. Namely, the process $\mu\rightarrow e\gamma$ 
induced at the one loop level via the Yukawa coupling $\lambda'_e$
is strongly suppressed due to the large mass of $H'_d$.
Furthermore, if the scale of mediation of SUSY breaking is larger
than the mass of the heavy Higgs doublets, which is the case for
$m_{3/2} \gtrsim 10\GeV (M/10^{16}\GeV)$, the couplings $\lambda'_e$
and $\lambda'_d$ induce off-diagonal entries in the squark and
slepton mass matrices through renormalization group running,
which in turn induce flavour violating processes in the quark and
lepton sector through quantum effects. However, these effects appear then at
the two loop level and are naturally suppressed. The most stringent bound
coming from $\mu\rightarrow e\gamma$ implies mild constraints
of the order $(\lambda_e'^\dag\lambda_e')_{21}\lesssim 0.1\, (m_\lstau/250\GeV)^2$,
whereas the $31$ and $32$ entries can be order one. Moreover, these effects can be
further suppressed by appropriate choices of the flavour structure of the
matrices  $\lambda'_e$ and $\lambda'_d$.
Therefore, in the scenario proposed in section \ref{Model}
it is expected that the stopped staus will decay not 
only into taus, but also into muons and electrons with
a large rate, while being consistent with all present constraints
on lepton flavour violation. In contrast, in scenarios with gravitino
LSP and stau NLSP the rates of lepton flavour violating stau decays
are predicted to be small~\cite{Hamaguchi:2004ne}. Therefore, 
the number of lepton flavor violating stau decays offers a sensitive probe 
to discriminate gravitino LSP scenarios
from the hidden fermion scenario discussed in section \ref{Model}.\bigskip

In section \ref{sec:Stable} we discussed the possibility that the gravitino and the hidden
fermion could be stable at cosmological timescales.  Nevertheless, the heavier
dark matter particle will eventually decay into the lightest and a tau-antitau
pair with a decay rate which is doubly suppressed by the Planck mass and by
the small coupling between the stau and the hidden fermion. Namely, when the
gravitino is lighter than the hidden fermion, the latter decays with a rate
which reads
\begin{equation}
	\Gamma(\hf \rightarrow \tau^+\tau^-\psi_{3/2})  =  
	\frac{|\lambda_{\lstau}|^2}{18432 \pi^3} \frac{m_\hf^9}
	     {M_P^2m_{3/2}^2m_\lstau^4}\;,
\end{equation}
where we have assumed $m_{3/2}\ll m_X$.  On the other hand, when the hidden
fermion is lighter than the gravitino, the decay rate approximately
reads~\cite{Buchmuller:2007ui}
\begin{equation}
  \Gamma(\psi_{3/2}\rightarrow \tau^+ \tau^- X)\simeq
  \frac{|\lambda_{\lstau}|^2}{92160\pi^3}\frac{m^7_{3/2}}{M_P^2 m_{\lstau}^4}
  \;.
\end{equation}
 
The electrons, positrons, gamma rays and neutrinos produced in the decay could
be in principle detected in cosmic-ray observations as an excess over the
expected astrophysical backgrounds. For dark matter particles with a mass
$\sim 100\GeV$, the decay products are observable if the lifetime is shorter
than $\sim 10^{26}\s$~\cite{Ibarra:2008jk}.  However, in our scenarios, the
heavier component of dark matter has a lifetime
\begin{equation}
  \tau(X\rightarrow \tau^+ \tau^- \psi_{3/2})\sim 
  10^{35}\s
  \left(\frac{|\lambda_{\lstau}|}{10^{-12}}\right)^{-2}
  \left(\frac{m_{X}}{100\GeV}\right)^{-9}
  \left(\frac{m_{3/2}}{10\GeV}\right)^2
  \left(\frac{m_{\lstau}}{250\GeV}\right)^{4}\;,
\end{equation}
\begin{equation}
  \tau(\psi_{3/2}\rightarrow \tau^+ \tau^- X)\sim 
  10^{38}\s
  \left(\frac{|\lambda_{\lstau}|}{10^{-12}}\right)^{-2}
  \left(\frac{m_{3/2}}{100\GeV}\right)^{-7}
  \left(\frac{m_{\lstau}}{250\GeV}\right)^4\;,
\end{equation}
which have been normalized to typical values of the masses and couplings which
yield a thermal history of the Universe incorporating thermal leptogenesis and
successful BBN ({\it cf.} Fig. 2). Therefore, the cosmic-ray fluxes typically
lie many orders of magnitude below the background, being completely
unobservable.

\section{Conclusions}
\label{Conclusions}

Cosmological scenarios where the observed matter-antimatter asymmetry is
generated by the supersymmetric thermal leptogenesis mechanism generically
fail to reproduce the observed abundances of primordial elements. In the
minimal scenario, the LSP must be a gravitino heavier than $\sim 5 \GeV$.
Therefore, if R-parity is conserved, the NLSP is very long lived, jeopardizing
the successful predictions of the standard BBN scenario. To solve this
conflict we have postulated the existence of a light hidden sector fermion
which couples very weakly to the NLSP. If the coupling is large enough, the
NLSP will decay dominantly into hidden sector fermions before the epoch of
primordial nucleosynthesis, avoiding all the nucleosynthesis constraints
altogether. We have analyzed the constraints on this coupling from dark matter
overproduction and from structure formation and we have found a wide window of
parameters where the cosmological history of the Universe can be consistent
with baryogenesis through thermal leptogenesis and with BBN. We have presented
two concrete models to illustrate the viability of the above mentioned
scenario.
Furthermore, we have discussed some experimental signatures at particle colliders
which can provide evidence for our mechanism and distinguish it from other
scenarios. Examples of these are the number of Higgs events and the lepton
flavour violation in late stau decays. Finally, we demonstrated that
the scenario is easily compatible with bounds from cosmic-ray observations.
        
\section*{Acknowledgements}
AI and CW would like to thank the IPMU for kind hospitality during the
completion of this work.  ADS gratefully thanks the warm hospitality of the
Physics Department at TUM where part of this work was done.  The work of MG
and AI was partially supported by the DFG cluster of excellence ``Origin and
Structure of the Universe.'' The work of ADS was supported in part by the INFN
``Bruno Rossi'' Fellowship, and in part by the U.S.~Department of Energy (DoE)
under contract No.~DE-FG02-05ER41360.

\section*{Note added}
The authors of Ref.~\cite{Cheung:2010qf}, which appeared simultaneously to
this one, independently considered a similar mechanism to accommodate BBN
bounds and thermal leptogenesis for a specific
situation where a stau LOSP can decay to a goldstino.


\begin{thebibliography}{99}

\bibitem{seesaw} 
P.~Minkowski,
Phys.\ Lett.\ B {\bf 67} (1977) 421;
M. Gell-Mann, P. Ramond and R. Slansky, \emph{Proceedings
of the Supergravity Stony Brook Workshop}, New York 1979, eds. P. Van
Nieuwenhuizen and D. Freedman; 
T. Yanagida, \emph{Proceedinds of the
Workshop on Unified Theories and Baryon Number in the Universe}, Tsukuba,
Japan 1979, eds. A. Sawada and A. Sugamoto; 
R. N. Mohapatra, G. Senjanovic, 
\textit{Phys.Rev.Lett.} \textbf{44} (1980)912, \textit{ibid.} \textit{%
Phys.Rev.} \textbf{D23} (1981) 165; 
S.~L.~Glashow, \emph{The Future Of Elementary Particle Physics},
\textit{In *Cargese 1979, Proceedings, Quarks and Leptons*, 687-713 and
Harvard Univ.Cambridge - HUTP-79-A059 (79,REC.DEC.) 40p};
%
J.~Schechter and J.~W.~F.~Valle,
Phys.\ Rev.\ D {\bf 22} (1980) 2227.
%
\bibitem{Fukugita:1986hr}
  M.~Fukugita and T.~Yanagida,
  Phys.\ Lett.\  B {\bf 174} (1986) 45.
%
\bibitem{Davidson:2002qv}
  S.~Davidson and A.~Ibarra,
  Phys.\ Lett.\  B {\bf 535} (2002) 25;
%
  W.~Buchmuller, P.~Di Bari and M.~Plumacher,
  Annals Phys.\  {\bf 315} (2005) 305.
%
\bibitem{Bolz:2000fu}
  M.~Bolz, A.~Brandenburg and W.~Buchm\"uller,
  Nucl.\ Phys.\  B {\bf 606} (2001) 518
  [Erratum-ibid.\  B {\bf 790} (2008) 336];
%
  J.~Pradler and F.~D.~Steffen,
  Phys.\ Rev.\  D {\bf 75}, 023509 (2007);
%
  V.~S.~Rychkov and A.~Strumia,
  Phys.\ Rev.\  D {\bf 75}, 075011 (2007).
%
\bibitem{Kawasaki:2004qu}
  M.~Kawasaki, K.~Kohri and T.~Moroi,
  Phys.\ Rev.\  D {\bf 71} (2005) 083502.

\bibitem{Cyburt:2009pg}
  R.~H.~Cyburt, J.~Ellis, B.~D.~Fields, F.~Luo, K.~A.~Olive and V.~C.~Spanos,
  JCAP {\bf 0910} (2009) 021.
%
\bibitem{Komatsu:2010fb}
  E.~Komatsu {\it et al.},
  arXiv:1001.4538 [astro-ph.CO].

\bibitem{Jedamzik:2006xz}
  K.~Jedamzik,
  Phys.\ Rev.\  D {\bf 74} (2006) 103509.




\bibitem{Pospelov:2006sc}
  M.~Pospelov,
  Phys.\ Rev.\ Lett.\  {\bf 98} (2007) 231301.
%
\bibitem{catalyzedBBN}
  K.~Kohri and F.~Takayama,
  Phys.\ Rev.\  D {\bf 76}, 063507 (2007);
%
  M.~Kaplinghat and A.~Rajaraman,
  Phys.\ Rev.\  D {\bf 74}, 103004 (2006);
  R.~H.~Cyburt {\it et al.}
  JCAP {\bf 0611}, 014 (2006);
  K.~Hamaguchi {\it et al.}
 Phys.\ Lett.\  B {\bf 650} (2007) 268;
%
  J.~Pradler and F.~D.~Steffen,
  Phys.\ Lett.\  B {\bf 666}, 181 (2008);
%
  M.~Pospelov, J.~Pradler and F.~D.~Steffen,
  JCAP {\bf 0811} (2008) 020.

\bibitem{Jedamzik:2009uy}
  K.~Jedamzik and M.~Pospelov,
  New J.\ Phys.\  {\bf 11} (2009) 105028.

%
\bibitem{Kanzaki:2006hm}
  T.~Kanzaki {\it et al.}
  Phys.\ Rev.\  D {\bf 75} (2007) 025011.
%
\bibitem{DiazCruz:2007fc}
  J.~L.~Diaz-Cruz, J.~R.~Ellis, K.~A.~Olive and Y.~Santoso,
  JHEP {\bf 0705} (2007) 003.
%
\bibitem{Buchmuller:2007ui}
  W.~Buchm\"uller, L.~Covi, K.~Hamaguchi, A.~Ibarra and T.~Yanagida,
  JHEP {\bf 0703}, 037 (2007).
%
\bibitem{left-right}
  M.~Ratz, K.~Schmidt-Hoberg and M.~W.~Winkler,
  JCAP {\bf 0810} (2008) 026.

\bibitem{Boubekeur:2010nt}
  L.~Boubekeur, K.~Y.~Choi, R.~R.~de Austri and O.~Vives,
  JCAP {\bf 1004} (2010) 005;

  J.~Pradler and F.~D.~Steffen,
  Nucl.\ Phys.\  B {\bf 809} (2009) 318.
%
\bibitem{Pradler:2006hh}
  J.~Pradler and F.~D.~Steffen,
  Phys.\ Lett.\  B {\bf 648} (2007) 224.

\bibitem{Asaka:2000ew}
  T.~Asaka and T.~Yanagida,
  Phys.\ Lett.\  B {\bf 494} (2000) 297.

%
\bibitem{Asaka:2000zh}
  T.~Asaka, K.~Hamaguchi and K.~Suzuki,
  Phys.\ Lett.\  B {\bf 490}, 136 (2000).

\bibitem{Hall:2009bx}
  L.~J.~Hall, K.~Jedamzik, J.~March-Russell and S.~M.~West,
  JHEP {\bf 1003} (2010) 080.

\bibitem{Ibarra:2008kn}
  A.~Ibarra, A.~Ringwald and C.~Weniger,
  JCAP {\bf 0901} (2009) 003.


\bibitem{Kolb:1990vq}
  E.~W.~Kolb and M.~S.~Turner,
  Front.\ Phys.\  {\bf 69} (1990) 1.

\bibitem{McDonald:2004eu}
  P.~McDonald {\it et al.}  [SDSS Collaboration],
  Astrophys.\ J.\ Suppl.\  {\bf 163} (2006) 80.

\bibitem{Strigari:2006jf}
  L.~E.~Strigari, M.~Kaplinghat and J.~S.~Bullock,
  Phys.\ Rev.\  D {\bf 75} (2007) 061303;
  F.~Borzumati, T.~Bringmann and P.~Ullio,
  Phys.\ Rev.\  D {\bf 77} (2008) 063514.

\bibitem{Boyarsky:2008xj}
  A.~Boyarsky, J.~Lesgourgues, O.~Ruchayskiy and M.~Viel,
  JCAP {\bf 0905} (2009) 012.

\bibitem{Viel:2004bf}
  M.~Viel, M.~G.~Haehnelt and V.~Springel,
  Mon.\ Not.\ Roy.\ Astron.\ Soc.\  {\bf 354}, 684 (2004).

\bibitem{Feng:2004mt}
  J.~L.~Feng, S.~Su and F.~Takayama,
  Phys.\ Rev.\  D {\bf 70}, 075019 (2004).

\bibitem{ArkaniHamed:2006mb}
  N.~Arkani-Hamed, A.~Delgado and G.~F.~Giudice,
  Nucl.\ Phys.\  B {\bf 741} (2006) 108.

\bibitem{Holdom:1985ag}
  B.~Holdom,
  Phys.\ Lett.\  B {\bf 166}, 196 (1986);
  B.~Holdom,
  Phys.\ Lett.\  B {\bf 259} (1991) 329.

\bibitem{delAguila:1988jz}
  F.~del Aguila, G.~D.~Coughlan and M.~Quiros,
  Nucl.\ Phys.\  B {\bf 307}, 633 (1988)
  [Erratum-ibid.\  B {\bf 312}, 751 (1989)].

\bibitem{Dienes:1996zr}
  K.~R.~Dienes, C.~F.~Kolda and J.~March-Russell,
  Nucl.\ Phys.\  B {\bf 492}, 104 (1997).

\bibitem{Abel:2008ai}
  S.~A.~Abel, M.~D.~Goodsell, J.~Jaeckel, V.~V.~Khoze and A.~Ringwald,
  JHEP {\bf 0807}, 124 (2008).

\bibitem{Goodsell:2009xc}
  M.~Goodsell, J.~Jaeckel, J.~Redondo and A.~Ringwald,
  JHEP {\bf 0911} (2009) 027.

\bibitem{Arvanitaki:2009hb}
  A.~Arvanitaki, N.~Craig, S.~Dimopoulos, S.~Dubovsky and J.~March-Russell,
  arXiv:0909.5440 [hep-ph].

\bibitem{Goodsell:2010ie}
  M.~Goodsell and A.~Ringwald,
  arXiv:1002.1840 [hep-th].

\bibitem{Giudice:1988yz}
  G.~F.~Giudice and A.~Masiero,
  Phys.\ Lett.\  B {\bf 206} (1988) 480.

\bibitem{Drees:1990yw}
  M.~Drees and X.~Tata,
  Phys.\ Lett.\  B {\bf 252} (1990) 695.


\bibitem{stopped-staus}
  W.~Buchmuller, K.~Hamaguchi, M.~Ratz and T.~Yanagida,
  Phys.\ Lett.\  B {\bf 588} (2004) 90;
%
  J.~L.~Feng and B.~T.~Smith,
  Phys.\ Rev.\  D {\bf 71} (2005) 015004
  [Erratum-ibid.\  D {\bf 71} (2005) 019904];
%
  K.~Hamaguchi, Y.~Kuno, T.~Nakaya and M.~M.~Nojiri,
  Phys.\ Rev.\  D {\bf 70} (2004) 115007;
%
  A.~Brandenburg, L.~Covi, K.~Hamaguchi, L.~Roszkowski and F.~D.~Steffen,
  Phys.\ Lett.\  B {\bf 617}, 99 (2005).

\bibitem{Hamaguchi:2004ne}
  K.~Hamaguchi and A.~Ibarra,
  JHEP {\bf 0502} (2005) 028
  [arXiv:hep-ph/0412229].

\bibitem{Ibarra:2008jk}
  A.~Ibarra and D.~Tran,
  JCAP {\bf 0902} (2009) 021;
  A.~Ibarra, D.~Tran and C.~Weniger,
  JCAP {\bf 1001} (2010) 009,
  Phys.\ Rev.\  D {\bf 81} (2010) 023529.

\bibitem{Cheung:2010qf}
  C.~Cheung, J.~Mardon, Y.~Nomura and J.~Thaler,
  arXiv:1004.4637 [hep-ph].



\end{thebibliography}
\end{document}